\documentclass[floatfix,twocolumn,aps,prb,showpacs,amsmath,amssymb,10pt]{revtex4-1}
\usepackage{amsmath}
\usepackage{cases}
\usepackage{amssymb}
\usepackage{amsfonts}
\usepackage{txfonts}
\usepackage[dvips]{graphicx}
\usepackage{color}

\begin{document}

\title{Electronic and transport properties in geometrically disordered graphene antidot lattices}
\author{Zheyong Fan$^{1,2}$}
\email{Corresponding author: brucenju@gmail.com}
\author{Andreas Uppstu$^{2}$}
\author{Ari Harju$^{2}$}
\affiliation{$^{1}$School of Mathematics and Physics, Bohai University, Jinzhou, 121000, China}
\affiliation{$^{2}$COMP Centre of Excellence, Department of Applied Physics, Aalto University, Helsinki, Finland}
\date{\today}

\begin{abstract}
A graphene antidot lattice, created by a regular perforation of a graphene sheet, can exhibit a considerable band gap required by many electronics devices. However, deviations from perfect periodicity are always present in real experimental setups and can destroy the band gap. Our numerical simulations, using an efficient linear-scaling quantum transport simulation method implemented on graphics processing units, show that disorder that destroys the band gap can give rise to a transport gap caused by Anderson localization. The size of the defect induced transport gap is found to be proportional to the radius of the antidots and inversely proportional to the square of the lattice periodicity. Furthermore, randomness in the positions of the antidots is found to be more detrimental than randomness in the antidot radius. The charge carrier mobilities are found to be very small compared to values found in pristine graphene, in accordance with recent experiments.
\end{abstract}

%-----------------------------------------------------------------------------80
\pacs{72.80.Vp, 72.15.Rn, 73.23.-b}
\maketitle
%-----------------------------------------------------------------------------80

\section{Introduction}

High quality graphene can reach a mobility of about $10^5$ cm$^2$/Vs even at room temperature \cite{bolotin2008}, making it a very promising material for future nanoelectronics. However, the absence of an appropriate band gap, needed in many applications in electronics and optoelectronics, limits its usability. There are various proposals for creating a band gap in graphene, including patterned hydrogenation \cite{balog2010} and the formation of a graphene antidot lattice (GAL) \cite{pedersen2008} (also known as graphene nanomesh \cite{bai2010}) by creating a pattern of nanometer-sized holes in graphene. The underlying mechanism of these methods is the generation of a periodic potential modulation in graphene, which induces a band structure transformation associated with a band gap opening. Electronic structure calculations indicate \cite{petersen2011, ouyang2011} that the band gap in GALs can be tuned by controlling the size, shape, and symmetry of the GAL unit cell.  The validity of these conclusions relies on the crucial assumption that the antidots form a perfectly ordered superlattice. While the self assembly on graphene could possibly be a promising route toward this \cite{paivi}, the effect of the deviation from the perfect periodicity is still an important question to be answered. This is especially important because according to the scaling theory of Anderson localization \cite{abrahams1979}, electrons in low-dimensional systems become more easily localized than in 3D materials. In a disordered low-dimensional system at low temperature, Anderson localization occurs when the system length exceeds the localization length, causing a transport gap that may not be easily distinguished from a gap in the band structure.

Experimentally produced GALs contain a significant amount of disorder \cite{eroms2009, kim2010, liang2010, zeng2012, giesbers2012, zhang2013, wang2013}, manifesting itself in fluctuations in both the antidot radius and location. Most experimental works have suggested that the observed gaps are more likely to be transport gaps rather than band gaps \cite{eroms2009,giesbers2012,zhang2013}, whereas in some studies the opposite conclusion has been reached \cite{kim2010, liang2010}. In this paper we study the localization properties of GALs, showing that only little disorder is needed for the localization length to become smaller than the usual size of experimental devices.

Previous theoretical works on the transport properties of GALs have mainly used the Landauer-B\"uttiker method combined with the recursive Green's function technique, which, due to the cubic-scaling of the computational effort with respect to the width of simulated system, is not suited to study localization properties in realistically sized samples. Thermoelectric properties of GALs in the quasi-1D limit and the ballistic regime have been studied using this method \cite{gunst2011,karamitaheri2011,yan2012}. Recent work by Power and Jauho \cite{power2014} showed that the transmission of an imperfect finite GAL connected to semi-infinite leads decays exponentially with respect to the length of the GAL. The conduction properties of GALs with anisotropic disorder have been studied by Pedersen \emph{et al.} \cite{pedersen2014}. Yuan \textit{et al.} \cite{yuan2013a,yuan2013b} have studied optical conductivity properties of GALs using an efficient numerical method based on Kubo formulas. Although their method can be used to study realistically sized samples, it cannot adequately handle localization effects, see Ref. [\onlinecite{fan2014cpc}] for discussion. Additionally, also the Dirac equation with a mass term has recently been applied to study electronic transport in GALs \cite{thomsen2014}. As such a model is scale-invariant, it is very suitable for studying large-sized systems, but it was found to be inaccurate for systems with significant amounts of disorder \cite{thomsen2014}.

In this work, we computationally study the electronic and transport properties of GALs in the presence of disorder using the tight-binding model and the linear-scaling real-space Kubo-Greenwood (RSKG) method \cite{mayou1988,mayou1995,roche1997,triozon2002}. This method is especially suitable for computing intrinsic properties of GALs, such as density of states, conductivity, localization length and charge carrier mobility. The simulations are performed with a code efficiently implemented on graphics processing units (GPUs) \cite{fan2014cpc}. By calculating the density of states and the conductivity at different length scales, we are able to give a detailed comparison between the band gap and the transport gap in disordered GALs.

Our simulations show that while the band gap of an antidot lattice vanishes in the presence of sufficiently strong disorder, it also gives rise to a transport gap through Anderson localization. Thus disorder effectively causes a transition from a band insulator into an Anderson insulator. The size of the transport gap depends on the spatial parameters of the antidot lattice, and our numerical results suggest a linear dependence on the antidot radius and inverse proportionality to the area of the unit cell of the antidot lattice. We also study charge carrier mobilities in disordered antidot lattices, and obtain realistic values compared with recent experimental results.

\begin{figure*}[hbtp]
\includegraphics[width=3.4in]{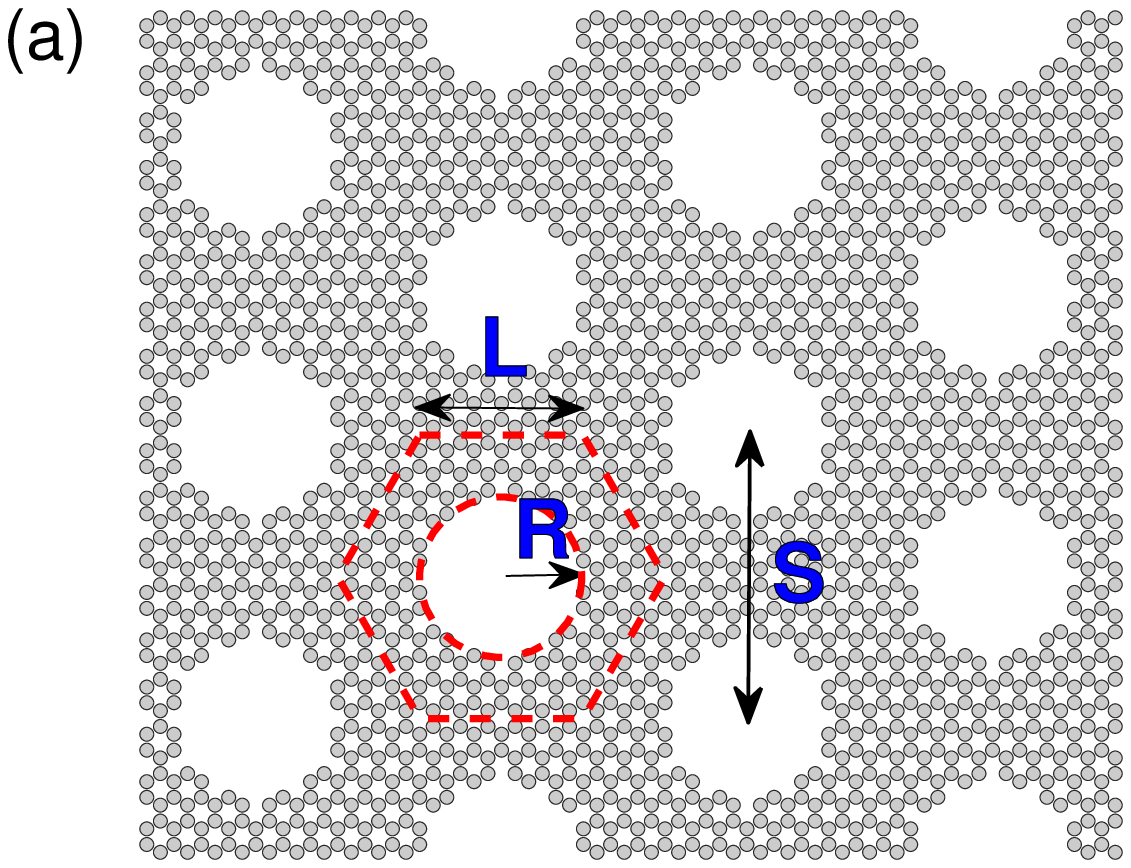}
\includegraphics[width=3.4in]{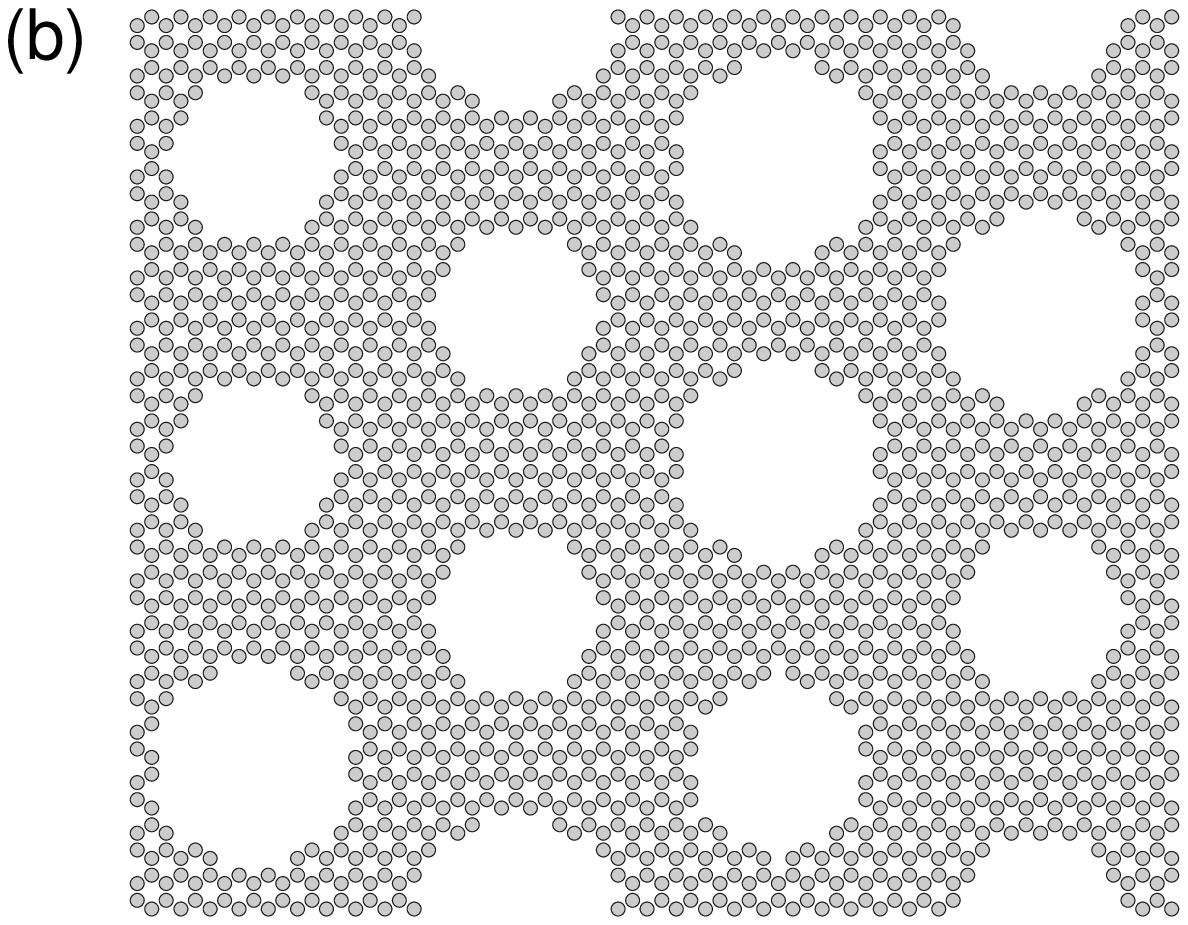}\\
\includegraphics[width=3.4in]{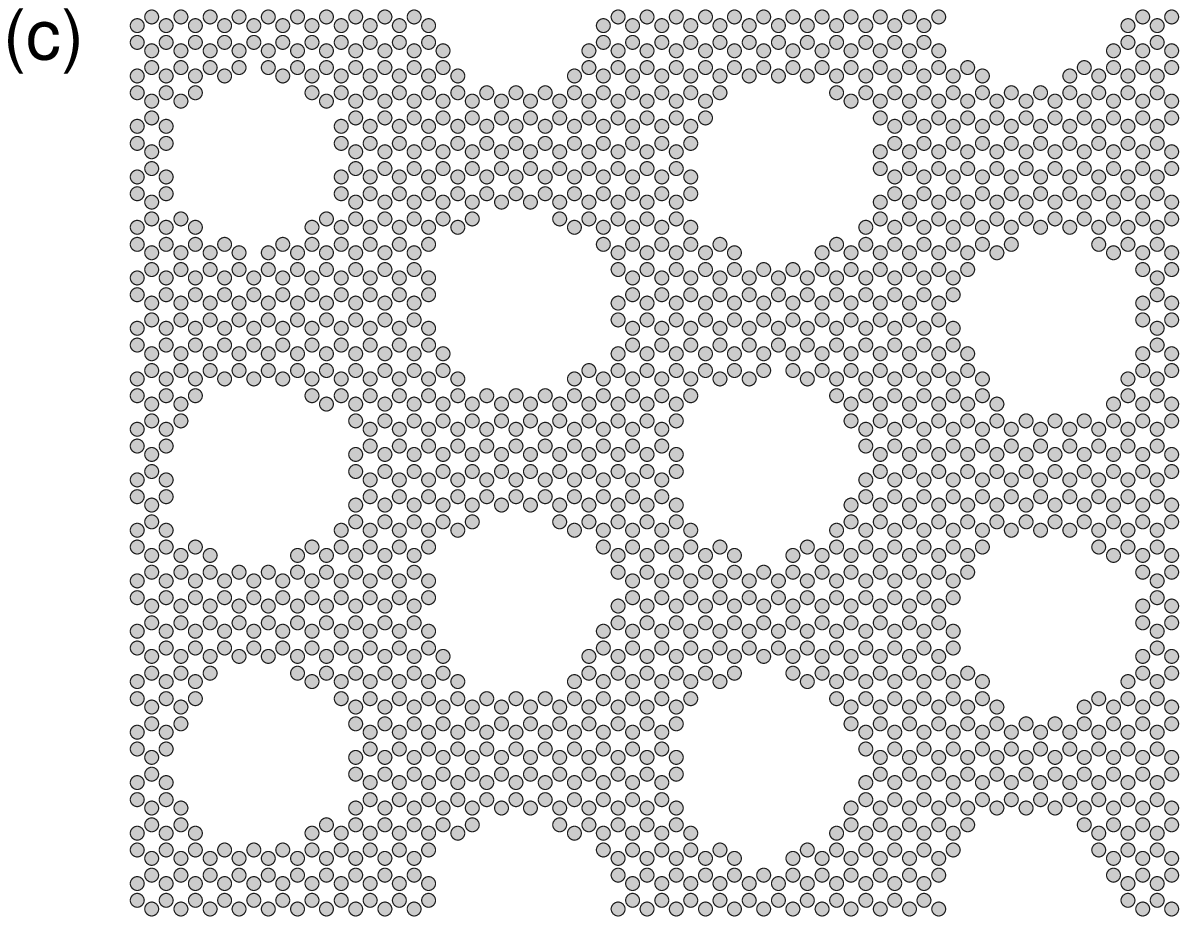}
\includegraphics[width=3.4in]{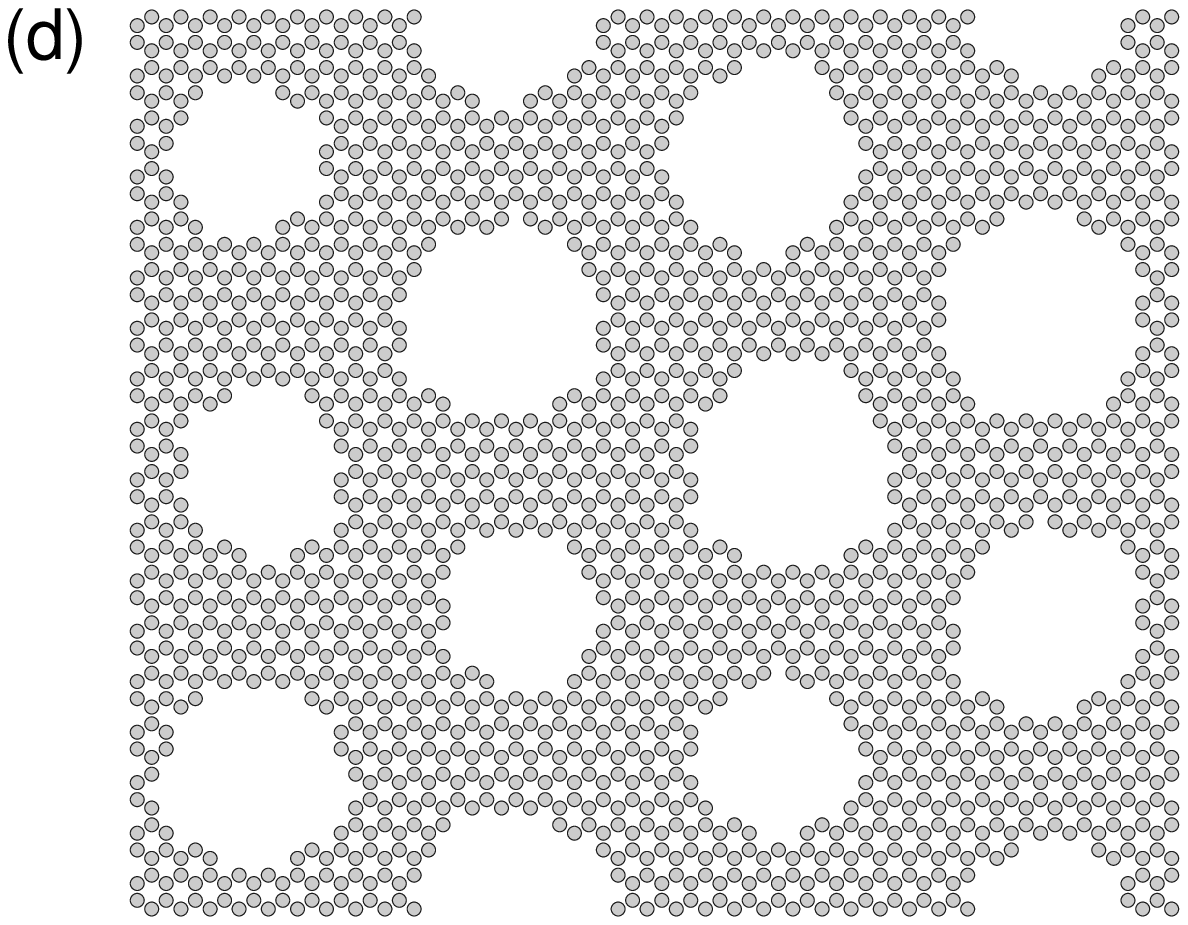}\\
\caption{(color online) Sketch of (6, 3)-GALs with (a) perfect periodicity, (b) radius disorder, (c) center disorder, and (d) mixed disorder. In (a), $S=\sqrt{3}L$ and $R$ represent the distances between the centers of two
adjacent holes (antidots) and the radius of the holes, respectively. Periodic boundary conditions are applied on both directions. Here, we have only shown relatively small systems for clarity, but in our simulations, we have considered much larger systems.}
\label{figure1}
\end{figure*}

\section{Models and Methods}
\label{section:models_and_methods}

\subsection{Models}
\label{section:models}

There can be various kinds of GALs, formed by different geometrical structures. To be specific, we follow Ref. [\onlinecite{pedersen2008}] and consider a GAL consisting of a triangular lattice of circular holes (antidots) in a graphene sheet. The radius of the holes is $R$ and the separation between the centers of nearest neighbor holes is $S$, see Fig.~ \ref{figure1} (a) for an illustration. We also require that in the pure antidot lattice the centers of the holes are always located in the middle of a hexagonal ring of the corresponding pristine graphene lattice. One can associate a hexagonal cell with side length $L=S/\sqrt{3}$ to each hole. A GAL with a side length $L$ and radius $R$ is labeled as $(L, R)$-GAL. Here, $L$ and $R$ are given in unit of $a$, the lattice constant of pristine graphene, which is set to be $0.246$ nm is this work. The hexagonal cell with side length $L$ contains $6L^2$ carbon atoms before the hole is created inside it. In order to match the triangular lattice of the GAL with the hexagonal lattice of pristine graphene, $L$ should be an integer. On the other hand, $R$ can be arbitrary, but one should note that for some values of $R$, there are carbon atoms with only a single neighboring atom. Such atoms are not likely to exist in real systems and are thus removed by hand. Also, carbon atoms with only two neighboring atoms are assumed to be passivated with hydrogen atoms. In this way, the electronic properties of the system close to half filling can be well described by the widely used nearest-neighbor $p_z$-orbital tight-binding Hamiltonian with a hopping parameter of $2.7$ eV.

There can be various kinds of disorder in GALs. In this work, we focus on disorders which are specific to GAL, leaving complications of coexistence of other conventional disorders found in graphene for a possible future study. In pristine GAL, the radii of the antidots are uniformly $R$ and the centers of the antidots form a perfect triangular lattice. Fluctuations of radii and centers can be regarded as radius and center disorder, respectively, which are collectively referred to as geometrical disorder \cite{yuan2013a,yuan2013b,power2014}.
We quantify the radius disorder by $\delta_R$ in such a way that the radii $R_i$ of the antidots take the following values with uniform probability:
\begin{equation}
\label{equation:radius_disorder}
R - \delta_R < R_i < R + \delta_R.
\end{equation}
Figure \ref{figure1}(b) is an illustration of the radius disorder. Similarly, we quantify the center disorder by $\delta_{xy}$ in such a way that the positions $(x_i, y_i)$ of the centers of antidots take the following values with uniform probability:
\begin{eqnarray}
\label{equation:radius_disorder}
&x_i^0 - \delta_{xy} < x_i < x_i^0 + \delta_{xy}, \\
&y_i^0 - \delta_{xy} < y_i < y_i^0 + \delta_{xy},
\end{eqnarray}
where $(x_i^0, y_i^0)$ are the positions of the centers of antidots in the pristine GAL. Figure \ref{figure1}(c) illustrates this disorder. While previous works \cite{yuan2013a,yuan2013b,power2014} studied these disorders separately, we also consider the situation where these two types of disorder coexist, as illustrated in Fig.~\ref{figure1}(d). Both $\delta_R$ and $\delta_{xy}$ are in units of $a$.

Since our focus is to study the 2D transport properties of GAL, we consider GALs with realistically large sizes. In all our calculations, we create GAL on a pristine graphene of size about 0.25 $\mu\text{m}^2$. We have tested that the results do not change upon further increasing the simulation size, suggesting that finite-size effects have mostly been eliminated.

\subsection{Methods}

We use the RSKG method \cite{mayou1988,mayou1995,roche1997,triozon2002} to study the quantum transport properties of GALs. In this method, the dc electrical conductivity as a function of energy $E$ and correlation time $t$ at zero temperature is given by
\begin{equation}
\label{equation:sigma}
\sigma(E, t) = e^2 \rho(E) \frac{d\Delta X^2(E, t)}{2d t}.
\end{equation}
Here, $\rho(E)$ is electronic density of states (with spin degeneracy taken into account) defined as
\begin{equation}
\label{equation:dos}
\rho(E) = \frac{2\textmd{Tr}[\delta(E - H)]}{\Omega},
\end{equation}
where $H$ is the Hamiltonian and $\Omega$ is the area of the system in 2D, which is taken to be the area of the pristine graphene sheet uniformly. The crucial quantity in the RSKG method is the mean square displacement
defined by
\begin{equation}
\label{equation:msd}
       X^2(E, t) = \frac{\textmd{Tr}\{[X, U(t)^{\dagger}] \delta(E - H)~ [X, U(t)] \}}
                           {\textmd{Tr}[\delta(E - H) ]},
\end{equation}
where $X$ is the position operator and $U(t) = \exp[-iHt/{\hbar}]$ is the time evolution operator. The first crucial step of achieving linear-scaling computation is to approximate the traces in Eq.~(\ref{equation:msd}) by using random vectors \cite{weisse2006} $|\phi\rangle$, $\textmd{Tr}[A] \approx  \langle \phi |A| \phi \rangle$, where $A$ is an arbitrary operator. To evaluate the numerator of Eq.~(\ref{equation:msd}) in a linear-scaling way, we evaluate the time-evolution $[X, U(t)]| \phi \rangle$ using a Chebyshev polynomial expansion. Last, we use the kernel polynomial method to achieve linear-scaling in the evaluation of the Dirac $\delta$ functions involved in both the density of states and the mean square displacement. All the calculations have been significantly accelerated by using GPUs and the detailed algorithms can be found in Ref.~[\onlinecite{fan2014cpc}].

One of the advantages of the RSKG method over other variants of Kubo-Greenwood formula-based numerical methods is that a definition of length (which can be regarded as the average propagating length of the electrons) is possible \cite{fan2014cpc,leconte2011,lherbier2012,uppstu2014,fan2014prb} in terms of the mean square displacement,
\begin{equation}
\label{equation:length}
  L(E, t) = 2 \sqrt{\Delta X^2 (E, t)}.
\end{equation}
This time-dependent length, different from the simulation cell length, provides a connection between the conductivity and the conductance in a quasi-1D geometry. In purely ballistic systems, the calculated conductance becomes time-independent (length-independent) in a short correlation time, and was found \cite{fan2014cpc} to be consistent with Landauer-B\"uttiker calculations except for numerical problems around Van Hove singularities. This definition was also found \cite{fan2014cpc,uppstu2014} to be usable in the localized regime, except for the extreme regime where the mean square displacement converges. The converged mean square displacement, on the other hand, provides a means for computing the localization length directly
using  \cite{uppstu2014,fan2014prb},
\begin{equation}
\label{equation:xi}
\xi(E) = \lim_{t\rightarrow \infty}\frac{\sqrt{\Delta X^2(E, t)}}{\pi}.
\end{equation}

\section{pristine graphene antidot lattices}

\begin{figure}
\includegraphics[width=\columnwidth]{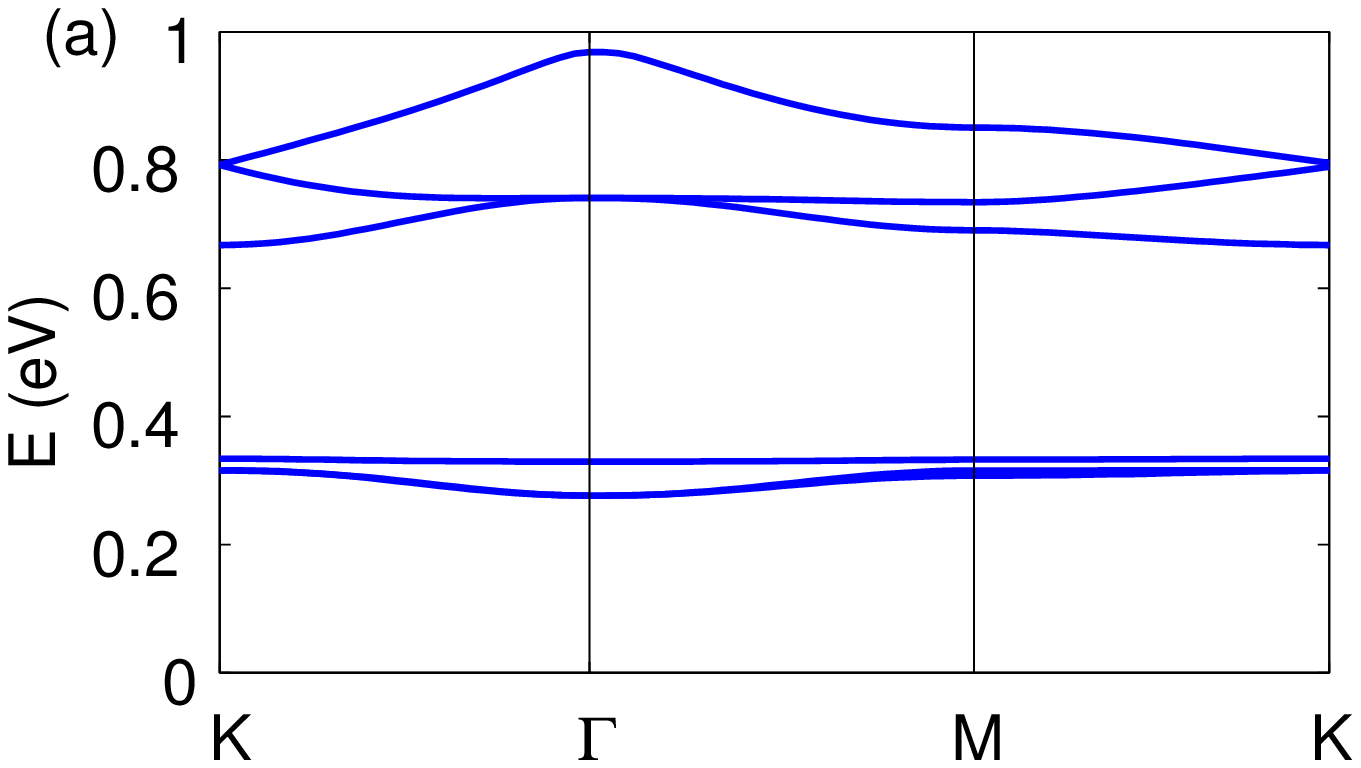}
\includegraphics[width=\columnwidth]{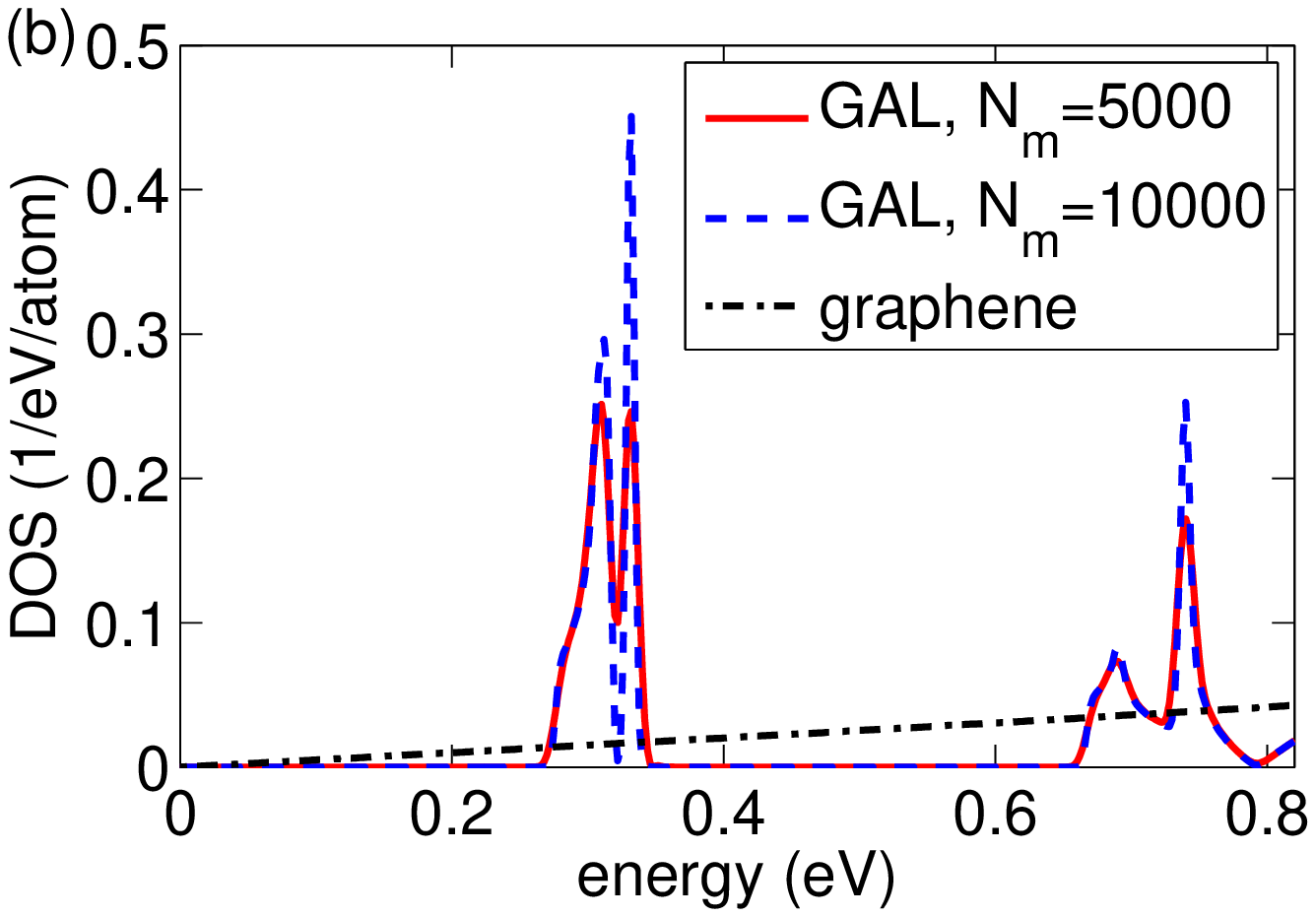}
\caption{(color online) (a) Band structure of a pristine (10, 6)-GAL. (b) Density of states of a pristine (10, 6)-GAL calculated using different energy resolutions, corresponding to different numbers of moments $N_m$, with that of pristine graphene also shown for comparison.}
\label{figure2}
\end{figure}

\begin{figure}
\includegraphics[width=0.9\columnwidth]{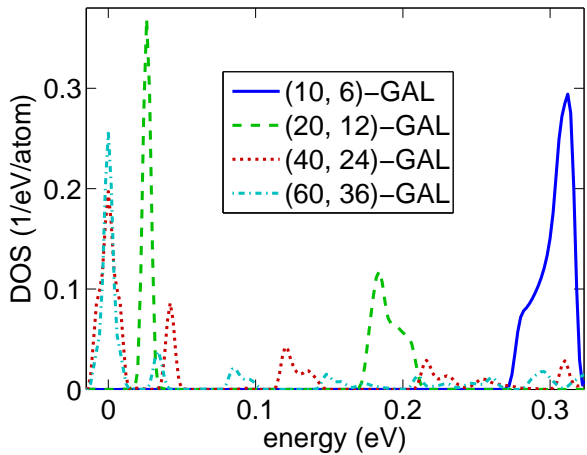}
\caption{(color online) Densities of states of pristine graphene antidot lattices.}
\label{figure:dos_10to60}
\end{figure}

We start presenting our results by first considering the electronic properties of perfect GALs, taking the (10, 6)-GAL as an example system. The band structure calculated using the nearest-neighbor tight-binding Hamiltonian is shown in Fig.~\ref{figure2}(a). Due to the particle-hole symmetry in the nearest-neighbor tight-binding model, only a relevant part on one side of the charge neutrality point (CNP) is shown for clarity. It is clear to see that there are two major band gaps in the range of 0 eV $<E<$ 1 eV.

While the density of states of the perfect system can be readily obtained from the band structure, a more efficient method, also applicable to disordered systems, is to use Eq.~(\ref{equation:dos}) and the linear-scaling techniques mentioned above. The results, however, depend on the energy resolution used, as shown in Fig.~\ref{figure2}(b). Here, $N_m$ is the number of Chebyshev moments used in the kernel polynomial method \cite{weisse2006}, which is associated with an energy resolution of order of $\Delta E/N_m$, where $\Delta E$ is the spectral width of the system. Our test shows that $N_m=10000$ is large enough to capture all the essential details of the band structure, while $N_m=5000$ is a little too small. We have used $N_m=10000$ in all the subsequent calculations.

We can similarly calculate the density of states of GALs with other values of $(L, R)$ using Eq.~(\ref{equation:dos}). Figure~\ref{figure:dos_10to60} shows the densities of states of four different  GALs. The values of $L$ are chosen to be 10, 20, 40, and 60, with the ratio $L/R$ being 5/3 in all cases. Since the band gap scales as\cite{pedersen2008} $R/L^2$, the band gap of a (20, 12)-GAL is much smaller than that of a (10, 6)-GAL around the CNP. For GALs with even larger constituent antidots, such as a (40, 24)-GAL and a (60, 36)-GAL, there is neither a band gap nor a Dirac point around the CNP. Instead, relatively flat bands appear around the CNP, resulting in a peak of density of states.

\section{Disordered graphene antidot lattices}

We now turn to discuss the effects of the geometrical disorder on the electronic and transport properties of GALs. We first study the effects of different types of the geometrical disorder. Three types of disorder are considered: radius disorder, center disorder, and a combination of these two.

\subsection{Effects of disorder type}

\begin{figure*}
\includegraphics[width=.68\columnwidth]{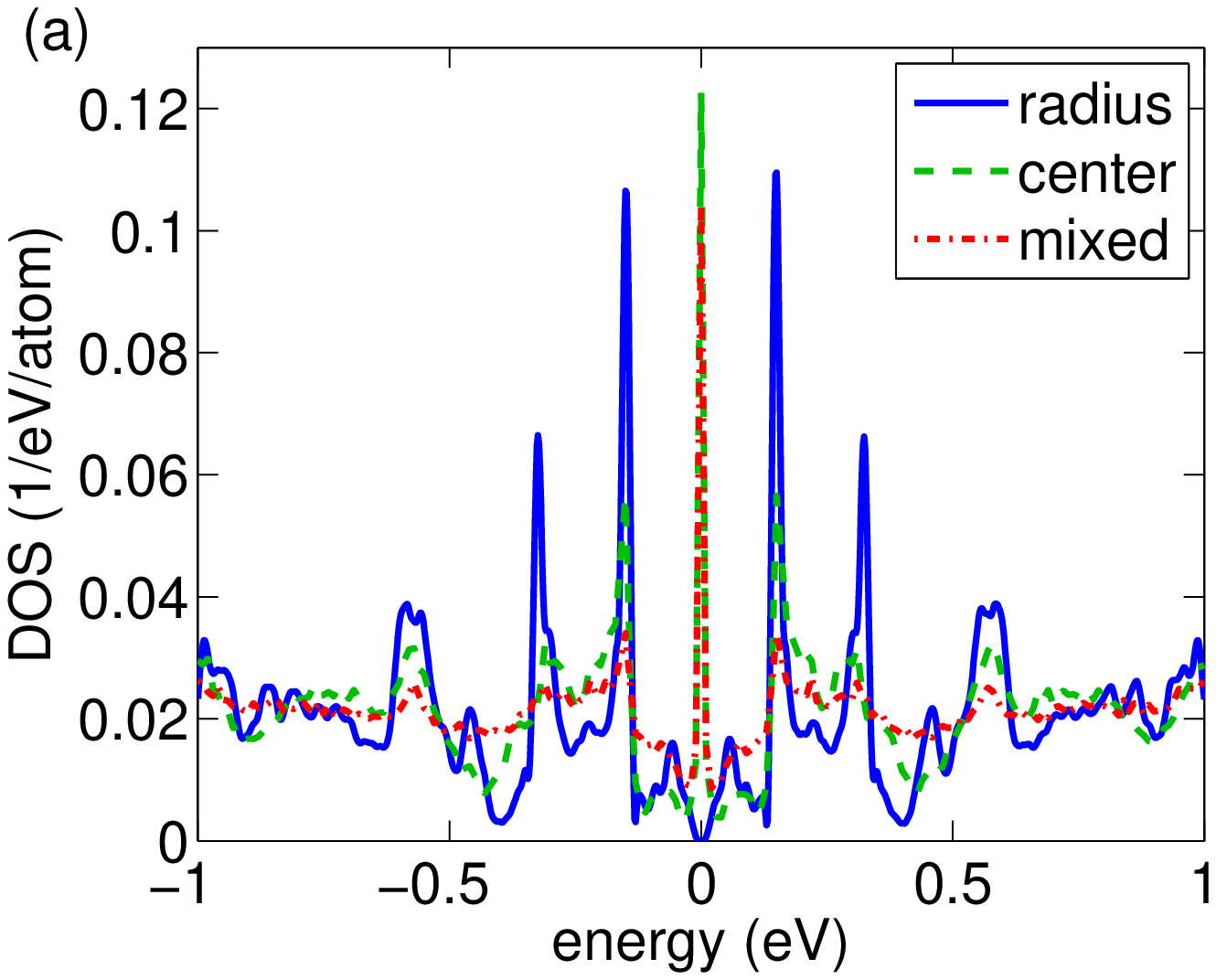}
\includegraphics[width=.68\columnwidth]{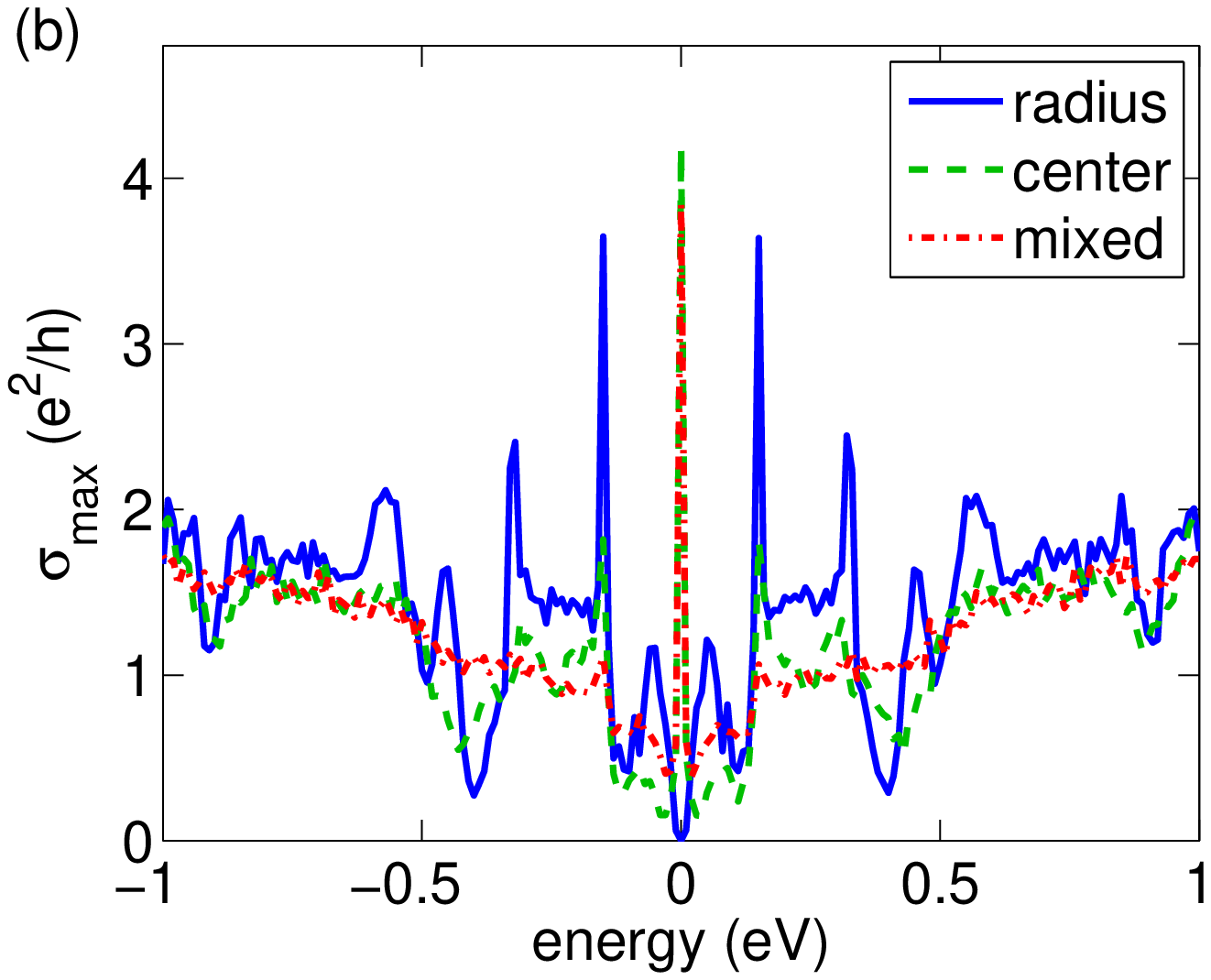}
\includegraphics[width=.68\columnwidth]{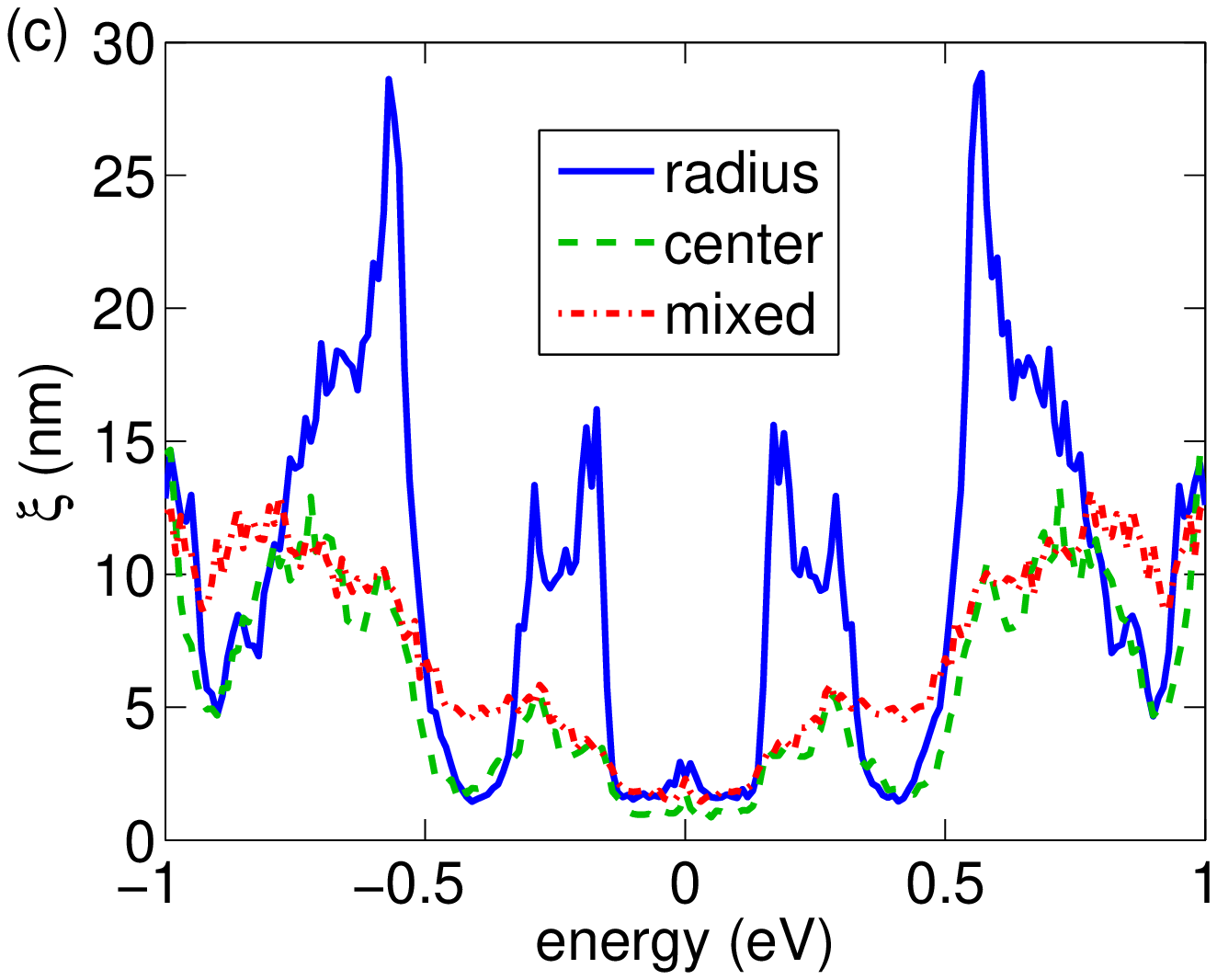}
\caption{(color online) (a) Density of states, (b) maximal conductivity, and (c) localization length as a function of energy for a (10, 6)-GAL with different types of geometrical disorder: radius disorder $\delta_R=0.5$ (solid lines), center disorder $\delta_{xy}=0.5$ (dashed lines), and mixed disorder $\delta_R=\delta_{xy}=0.5$ (dot-dashed lines).}
\label{figure:type}
\end{figure*}

For the different disorders, we again use the (10, 6)-GAL as a test case. Figure \ref{figure:type}(a) shows the densities of states of these systems. It is clear that the original band gap found in a pristine (10, 6)-GAL vanishes in the presence of each kind of disorder. An apparent difference between radius and center disorder is that there is a peak in the density of states at the CNP in the presence of the latter while it is absent in the presence of the former. This difference may be explained by the preservation of the rotational symmetry in the presence of pure radius disorder [cf. Fig.~\ref{figure1}(b)]. The center disorder, on the other hand, results in  antidots with irregular edges and breaks the rotational symmetry [Fig.~\ref{figure1}(c)]. The atoms at the irregular edges behave similarly as resonant scatterers, such as vacancies, whose presence also results in a sharp peak of density of states at the CNP \cite{fan2014cpc,fan2014prb}.

Transport properties can also be efficiently obtained by the RSKG method. In the presence of disorder, the diffusive (ohmic) transport is mainly characterized by the semiclassical conductivity, which, in the RSKG method is commonly defined to be the maximal conductivity over the correlation time at each energy. Figure~\ref{figure:type}(b) shows the maximal conductivities in a (10, 6)-GAL with different types of geometrical disorder. These exhibit a similar energy dependence as the densities of states. However, it should be noted that the sharp peaks in the maximal conductivity at the CNP in the presence of center and mixed disorder cannot be simply interpreted as the semiclassical conductivity. This is similar to the case of graphene with vacancy disorder \cite{fan2014prb}, as also the zero energy peak in the density of states plays a major role at that point.

No matter how well the semiclassical conductivity is approximated by the maximal conductivity, it is not a quantity that can be measured in samples with a fixed size. The reason is that localization takes place when the sample length exceeds the mean free path. At low temperature, the conductivity becomes vanishingly small when the sample length is several times larger than the localization length. Interestingly, the calculated localization lengths [\textit{via} Eq.~(\ref{equation:xi})] are found to be only of the order of 10 nm [Fig.~\ref{figure:type}(c)]. These localization lengths are comparable to those which can be extracted from the exponential decays of the transmissions in a (7, 3)-GAL with similar geometrical disorder studied by Power and Jauho \cite{power2014}. It can also be seen that in the largest part of the energy range tested here, the charge carriers in a GAL with center disorder become more easily localized than those in a GAL with radius disorder.

\subsection{Scaling of the transport properties with respect to the GAL parameters}

Above, the GAL lattice was, apart from the disorder, fixed to a certain geometry. Next, we keep the distance between the antidot centers fixed and study how changing the antidot radius affects the electronic and transport properties of the system. Both kinds of disorder are present in our simulations, as the results might then be best comparable with experiments. For simplicity, we first fix the disorder strength to be $\delta_R = \delta_{xy}$ = 1.0.

\begin{figure*}
\includegraphics[width=.68\columnwidth]{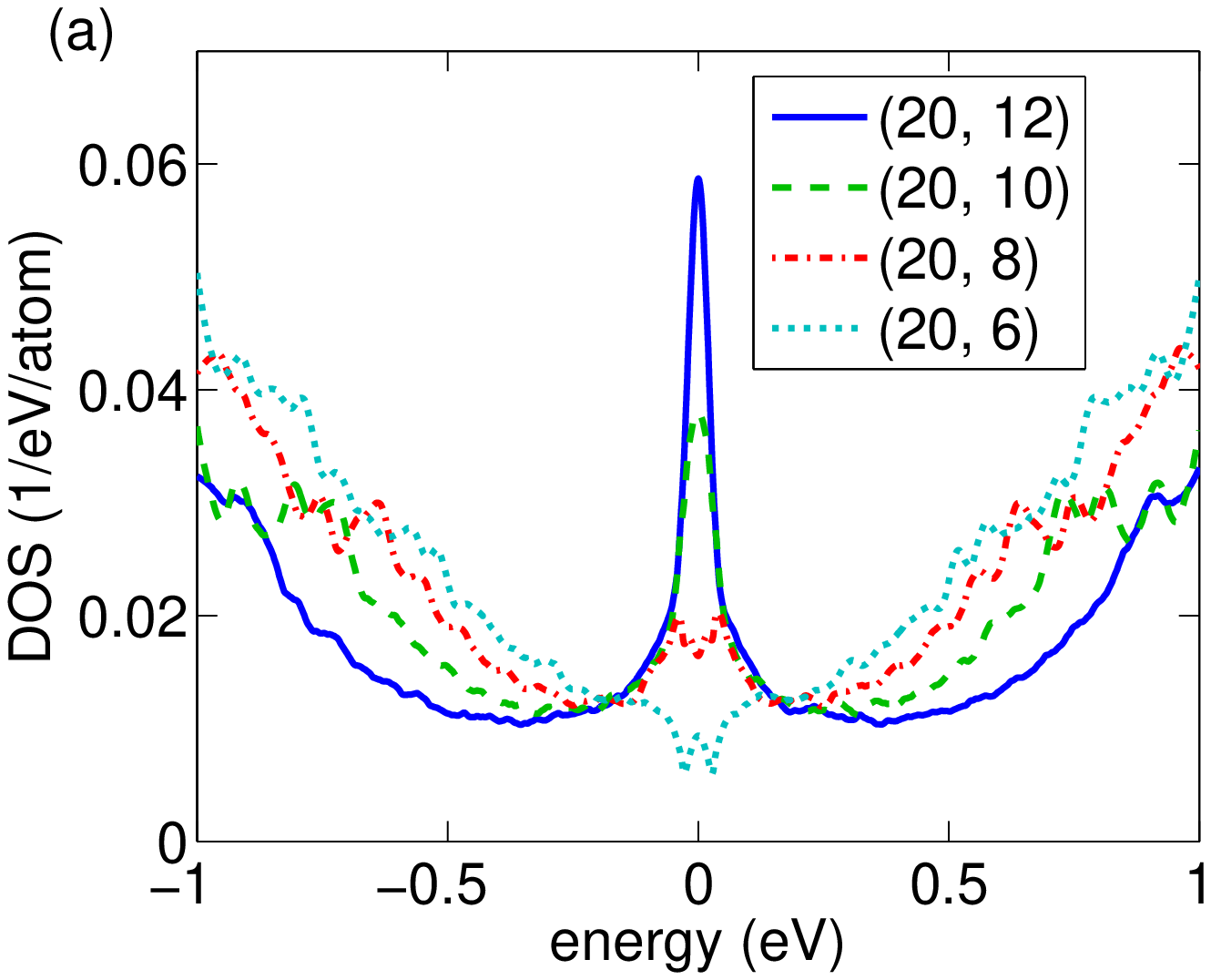}
\includegraphics[width=.68\columnwidth]{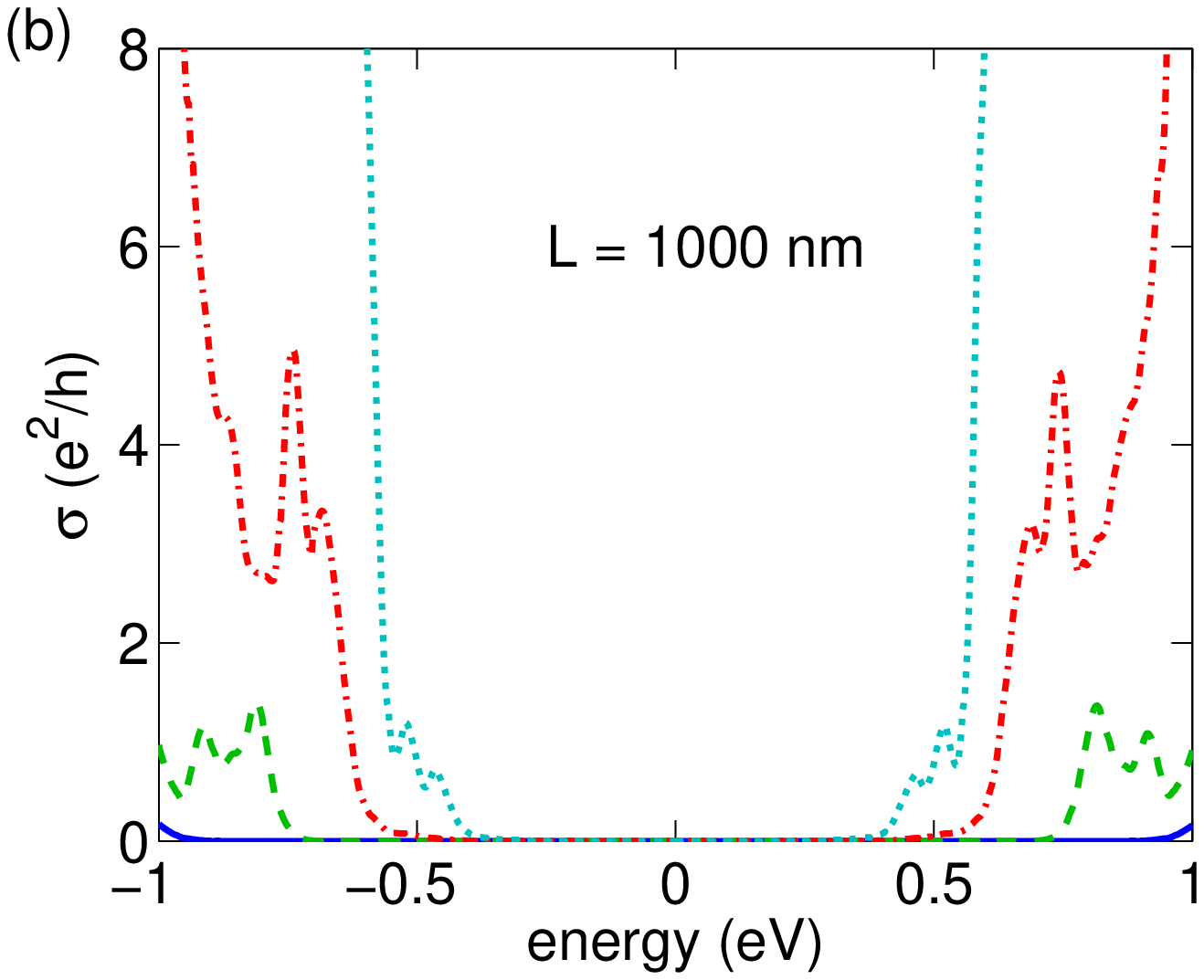}
\includegraphics[width=.68\columnwidth]{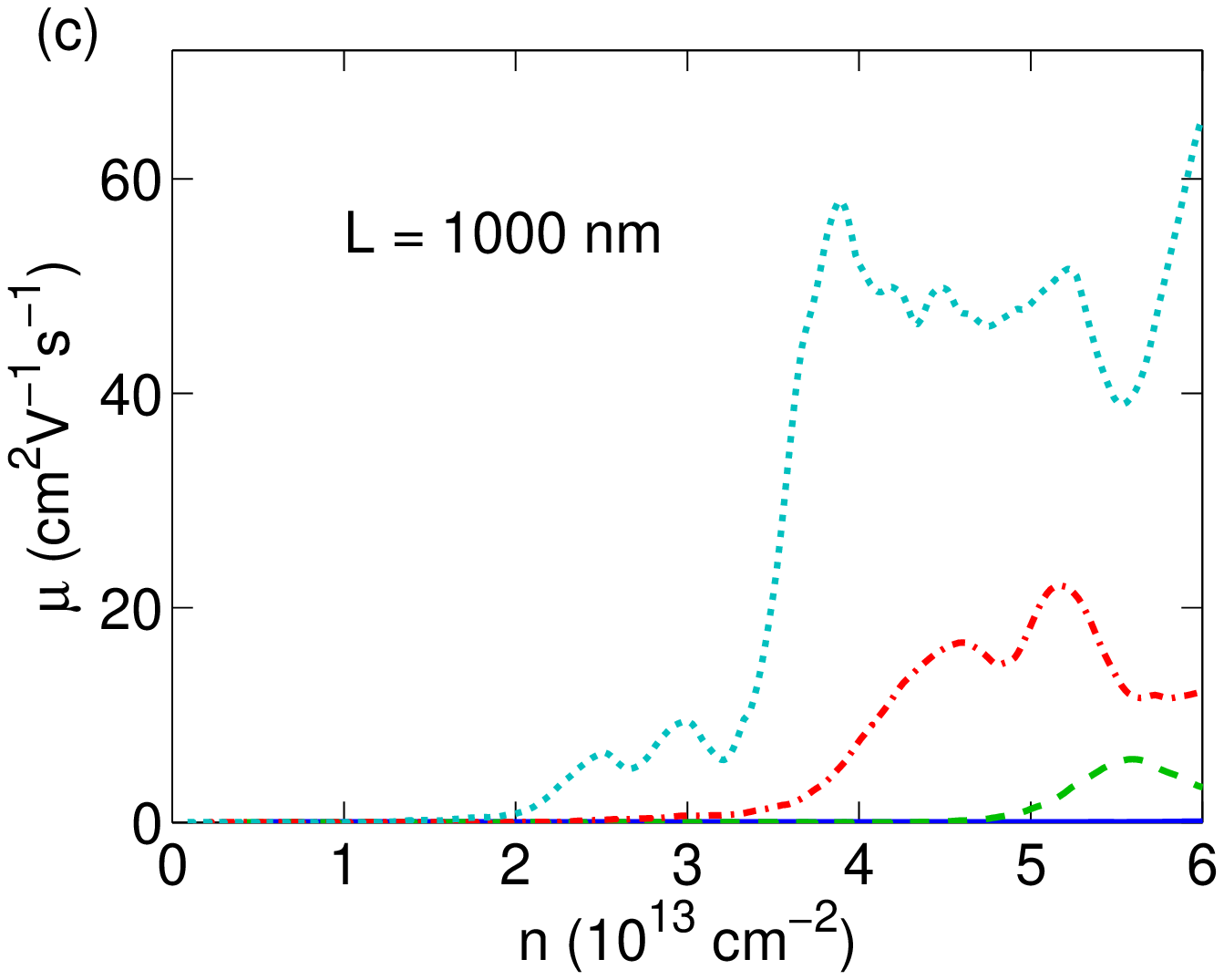}
\caption{(color online) (a) Density of states as a function of energy, (b) conductivity at a propagation length of 1000 nm as a function of energy and (c) mobility at the same length as a function of carrier concentration for a (20, 12)-GAL (solid line), a (20, 10)-GAL (dashed line), a (20, 8)-GAL (dot-dashed line), and a (20, 6)-GAL (dotted line) with the same mixed geometrical disorder of $\delta_R = \delta_{xy}=1.0$.}
\label{figure:scale_1}
\end{figure*}

% Ari
The density of states depends strongly on the antidot radius, as shown in
Fig.~\ref{figure:scale_1}(a). When the antidot radius decreases, the
zero-energy peak in the density of states becomes lower and the values
at finite energies become larger. The large peaks at the CNP do not,
however, contribute to the transport properties of the system, as the
peaks correspond to localized states. This can be seen in
Fig.~\ref{figure:scale_1}(b), which shows the conductivity at a fixed propagation length of 1000 nm. In the cases where
the propagation length of the simulation has not reached this value,
an extrapolation has been performed to reach the desired system size.
All the peaks around the CNP are in a regime where the conductivity
is low. One should note that the conductivity of the GAL with the
largest antidots is nearly zero for all the energies shown, having
non-zero contributions only around energies $\pm 1$ eV.
Furthermore, the figure shows that the effect of
decreasing antidot radius on the
conductivity is even more dramatic than on the density of states.
The transport gap of around 2 eV in the case with the largest
value of antidot radius reduces to around half of that as the antidot
radius is halved. This suggests a linear dependence of the gap on the
antidot radius. The scaling is analyzed in more detail below.

In order to make more direct comparisons with experiments, one can
extract the mobility $\mu$ from the conductivity using
$$\sigma(E, n) = e n(E) \mu(E,n) \ ,$$ where the carrier concentration
$n$ is calculated from the density of states as $n(E) =
\int_0^{E}\rho(E') dE'$ and the zero of the energy corresponds to the
CNP. The mobility obtained in this way is presented in Fig.~\ref{figure:scale_1}(c).
For the largest antidot radius, the mobility remains close to zero for
the whole range of charge carrier densities shown.
For smaller radius sizes, there is a clear increase of mobility with decreasing
antidot radius at finite charge carrier
densities. This would suggest that a GAL with a small antidot radius
would be beneficial for electronic{\bf s} applications that require a large
mobility outside the gap. It is also clear that the mobility of GAL is rather
low even when the charge carrier density is fairly large and the dot radius is small.
This agrees very well with experimental findings \cite{zhang2013}.

\begin{figure*}[bht]
\includegraphics[width=.68\columnwidth]{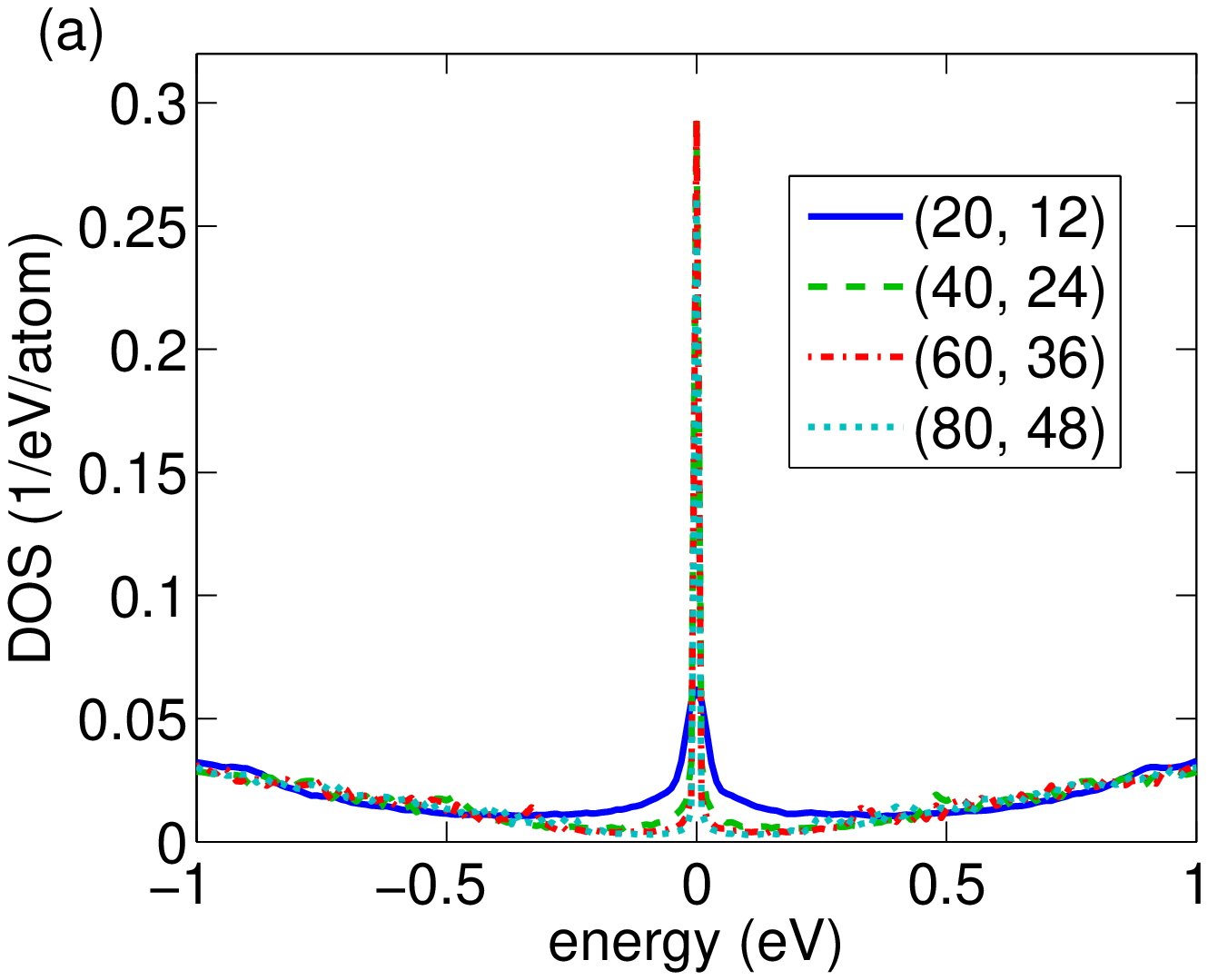}
\includegraphics[width=.68\columnwidth]{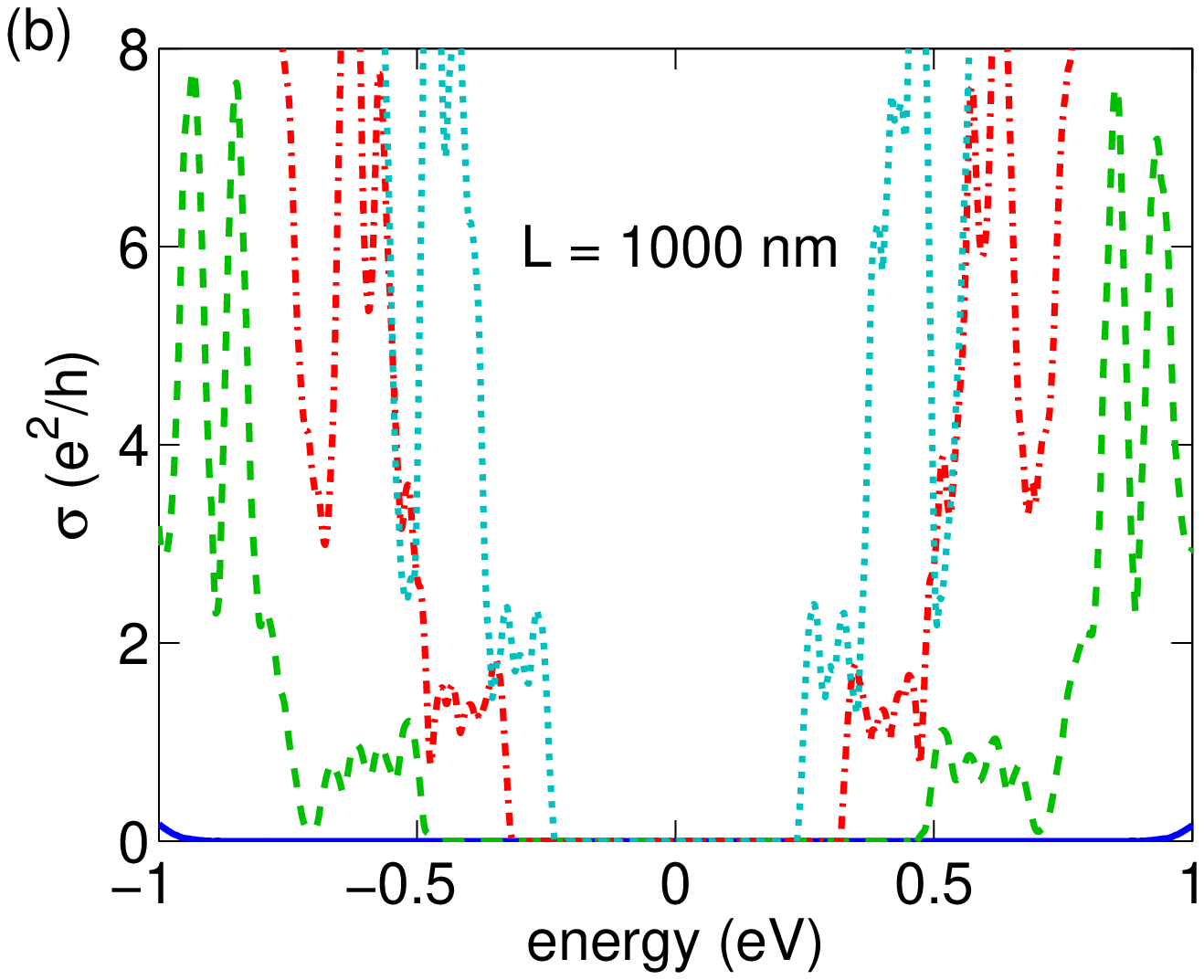}
\includegraphics[width=.68\columnwidth]{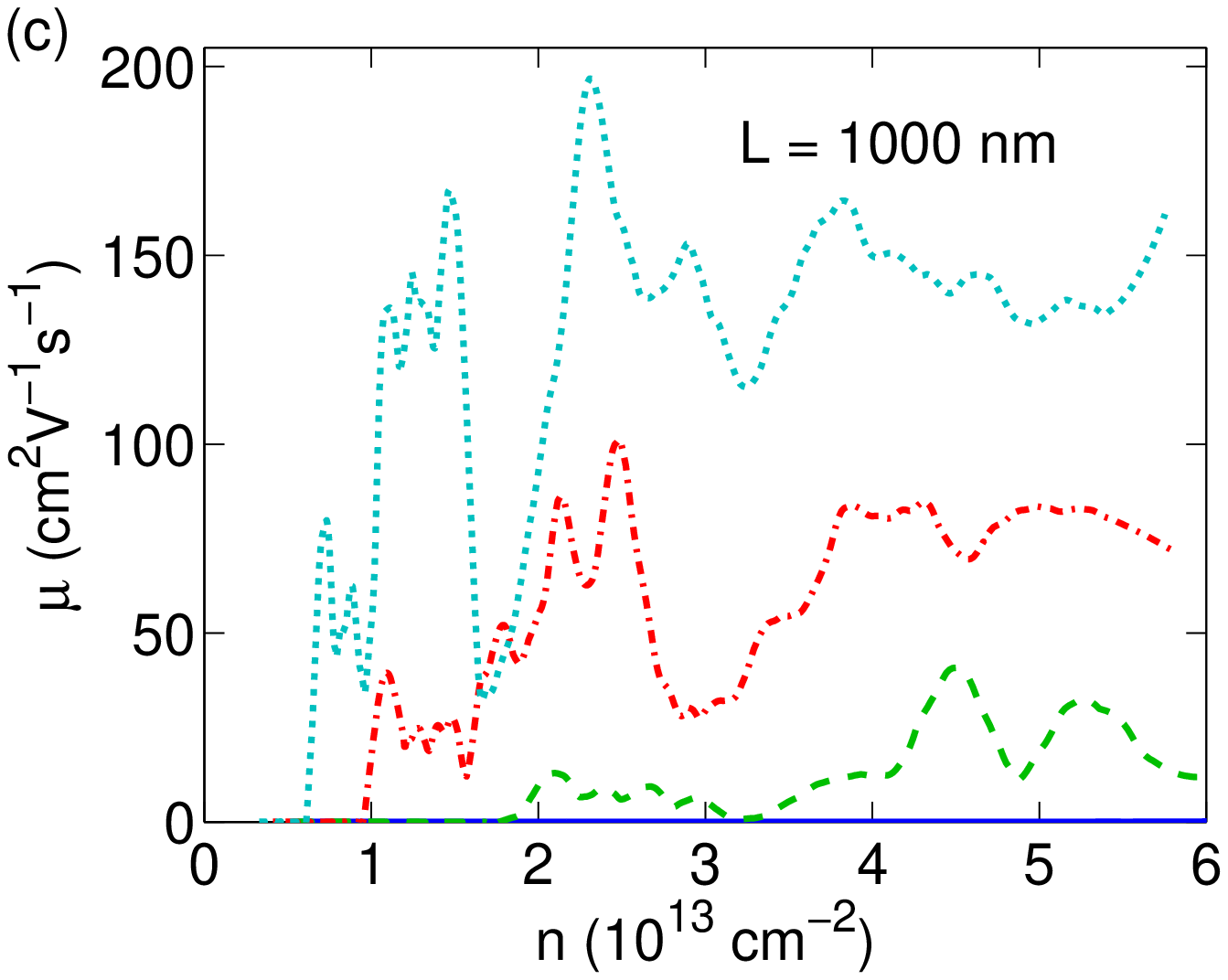}
\caption{(color online) (a) Density of states, (b) conductivity at a
  propagation length of 1000 nm as a function of energy and (c)
  mobility at the same length as a function of carrier concentration
  for a (20, 12)-GAL (solid line), a (40, 24)-GAL (dashed line), a
  (60, 36)-GAL (dot-dashed line), and a (80, 48)-GAL (dotted line)
  with the same mixed geometrical disorder of $\delta_R =\delta_{xy}=1.0$. }
\label{figure:scale_2}
\end{figure*}

% Ari
In the analysis above, the distance between the antidot centers was
fixed and only the antidot radius was allowed to change. Next, we vary
both the antidot radius and distance, keeping the ratio of these two
constant. We start with the density of states, shown in
Fig.~\ref{figure:scale_2}(a). All the system sizes shown have a very
sharp peak at the CNP, with the peak becoming even sharper with
larger lattice parameters. Apart from that, however, the
density of states of the different lattices shows now more similar
behavior than in the case above where only the antidot radius was varied.
Furthermore, like above, the states at the CNP region are localized as can be seen from
Fig.~\ref{figure:scale_2}(b) that shows the corresponding conductivities.
This time the larger lattices show irregular plateaus outside the transport gap.
This is possibly caused by the more complicated
band structures, which is also reflected in the density of states of the
pure systems shown in Fig.~\ref{figure:dos_10to60}. The gap region is
very clear with all lattice sizes, and the gaps become smaller as the lattice
parameters increase.

Again, the most direct comparison with the experiments can be done
using the mobility, extracted similarly as in the previous case of
varying antidot radius.  The mobility, shown in
Fig.~\ref{figure:scale_2}(c), is still rather low but larger than in
Fig.~\ref{figure:scale_1}(c) above.  The mobility at finite charge
carrier densities increases rather linearly as a function of lattice
size. This can be used to obtain extrapolated values for experimentally
relevant system sizes.  The experimental results of
Ref.~[\onlinecite{zhang2013}] were obtained for a GAL with similar
$R/L$ ratio, but with an around four times larger antidot lattice
size. A room temperature mobility of 750 cm$^2$/(V s) was measured,
which very interestingly agrees rather well with an approximation
based on a linear extrapolation of our simulation data.

\begin{figure}[h]
\includegraphics[width=.8\columnwidth]{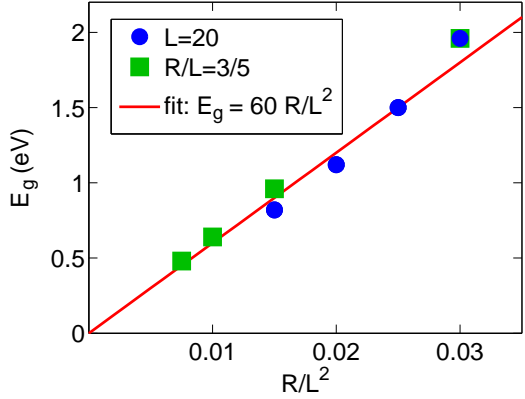}
\caption{(color online) Transport gap as a function of the antidot radius divided by the
  lattice size squared in GALs with the same mixed geometrical disorder of $\delta_R =
  \delta_{xy}=1.0$. The circles correspond to a (20, 12)-GAL, a (20, 10)-GAL, a (20, 8)-GAL, and a (20, 6)-GAL as studied in Fig.~\ref{figure:scale_1}. The squares correspond to a (20, 12)-GAL, a (40, 24)-GAL, a (60, 36)-GAL, and a (80, 48)-GAL as studied in Fig.~\ref{figure:scale_2}. The line represents a linear fitting of the data with $E_g=60R/L^2$. Both $L$ and $R$ are given in units of $a=0.246$ nm.}
\label{figure:E_g}
\end{figure}

% Ari
The results above suggest a scaling of the transport gap, which we now analyze in more detail. We define the gap as the energy range where
conductivity is below $0.1e^2/h$. Although this definition of a transport gap is
somewhat arbitrary, it does not affect our conclusions.  The gap
 for the conductivity data
presented in Figs.~\ref{figure:scale_1}(b)
and~\ref{figure:scale_2}(b) is shown in Fig.~\ref{figure:E_g}.  It can be seen that the transport gap is
indeed linearly proportional to the antidot radius. Furthermore, the
gap is inversely dependent on the square of the lattice constant of
the GAL. Surprisingly, the zero-temperature transport gap of disordered GALs scales thus similary as the band gap of pristine GALs \cite{pedersen2008}.

\begin{figure*}[hbt]
\includegraphics[width=.68\columnwidth]{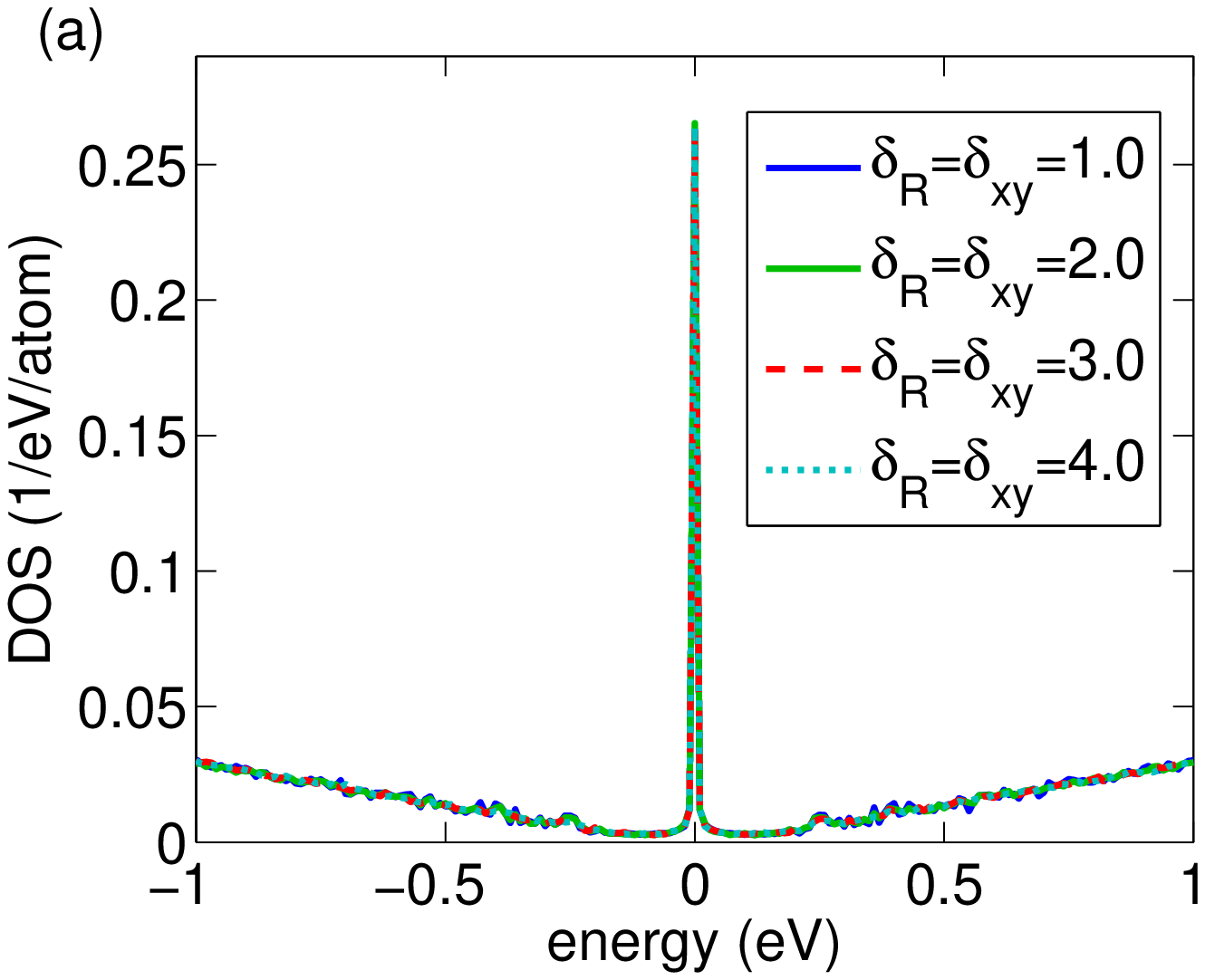}
\includegraphics[width=.68\columnwidth]{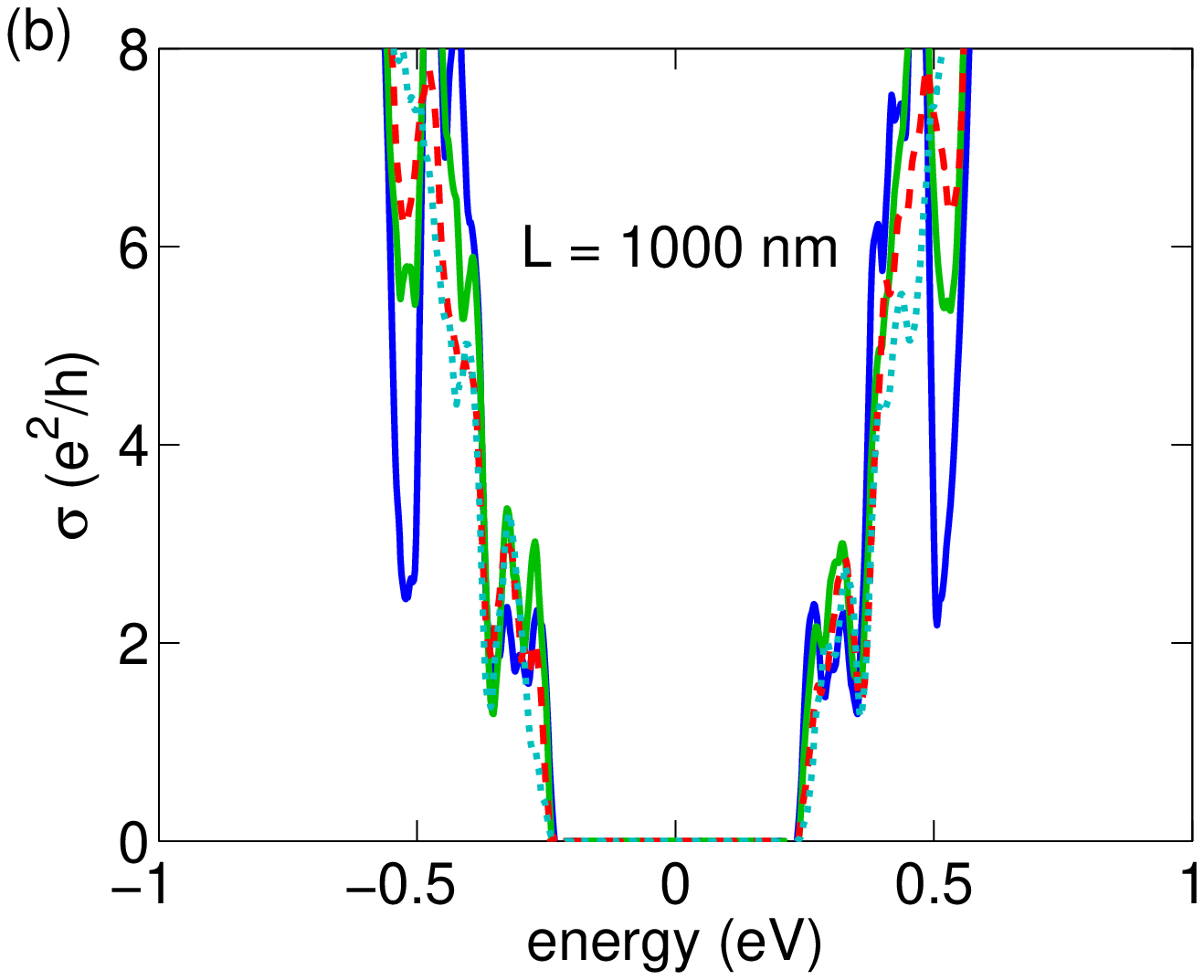}
\includegraphics[width=.68\columnwidth]{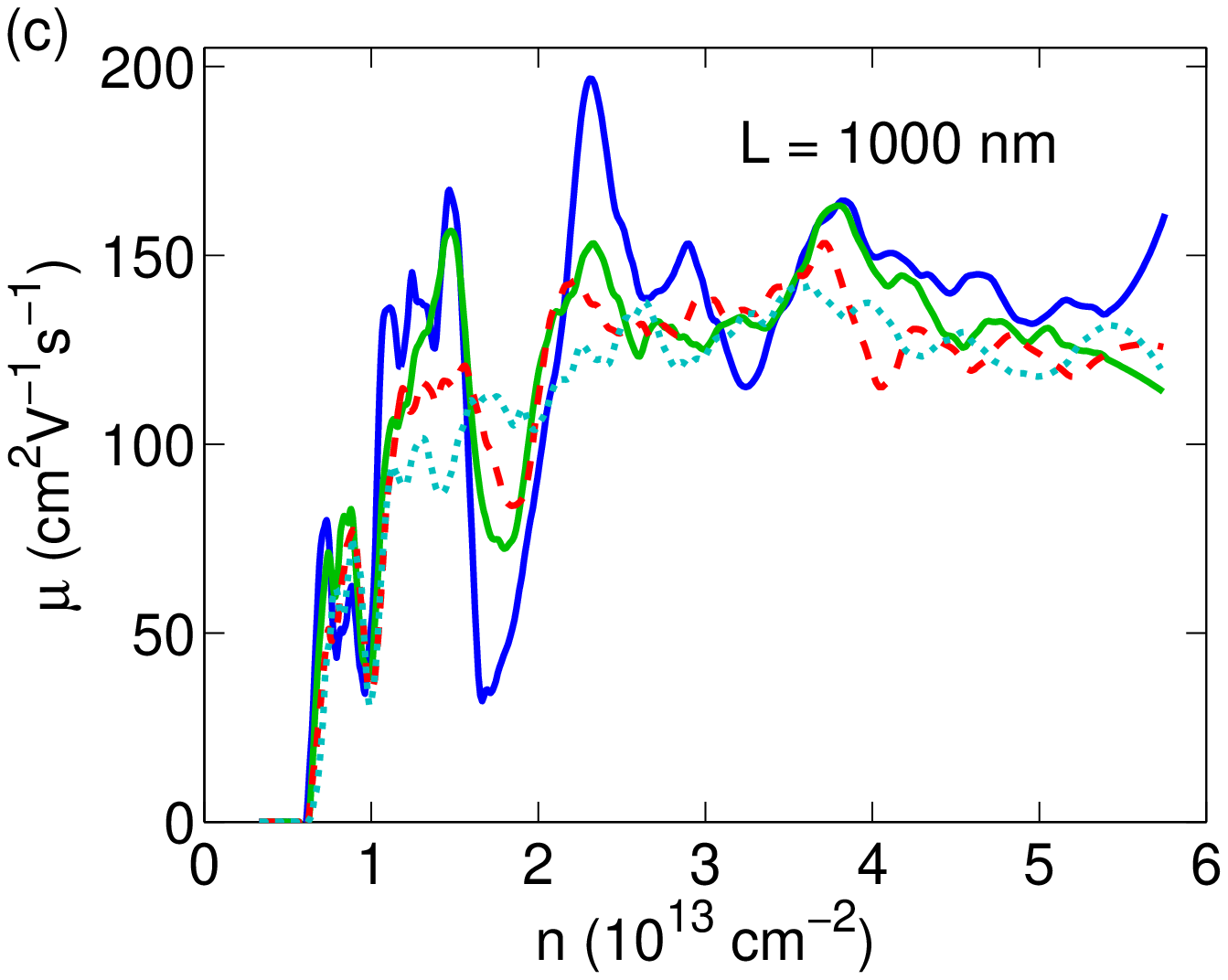}
\caption{(color online) (a) Density of states, (b) conductivity at a propagation length of 1000 nm as a function of energy and (c) mobility at the same length as a function of carrier concentration for a (80, 48)-GAL with different strengths of the mixed-type disorder $\delta_R=\delta_{xy}$.}
\label{figure:strength}
\end{figure*}

% Ari
The results above were obtained for GALs with a fixed amount of disorder, and by varying the lattice size and the antidot radius. We now study the
effects of varying the disorder strength in a fixed lattice geometry, using a (80, 48)-GAL as a test system. We start by comparing the densities of states, shown in
Fig.~\ref{figure:strength}(a) and obtained with the disorder strengths of
$\delta_R=\delta_{xy}=$ 1.0, 2.0, 3.0, and 4.0. The
results for all of these values of disorder strength are very similar in the
whole energy range, suggesting that arbitrary shifts of the GAL centers
by a single lattice constant of pristine graphene is sufficient to
cause significant disorder effects in the density of states.
Furthermore, the conductivity and mobility, shown in
Figs.~\ref{figure:strength}(b) and (c), indicate that the transport
properties of the disordered GALs are also very
similar. Although in the presence of less disorder
there are more fluctuations in the conductivity, the
transport gaps and mobilities are strikingly similar in all cases.
The results indicate that although even a small amount of disorder
in GALs is enough to cause a transport gap, the transport properties
of GALs are not especially sensitive to the disorder strength. This
highlights the importance of taking the disorder into account in
modeling GALs.  Furthermore, this indicates that the correspondence between our
theoretical predictions and the experimental measurements presented in Ref.~[\onlinecite{zhang2013}] is
likely not a coincidence, although the amount of disorder in the experimental setup
is not easy to extract.

\section{Conclusions}
We have simulated intrinsic transport properties of graphene antidot lattices using the Kubo-Greenwood formalism.  We have shown that geometrical disorder, which is associated with deviation from the perfect superlattice structure, easily causes the graphene antidot lattice to undergo a transition from a band insulator to an Anderson insulator. Our results support the conclusions of several experimental studies, \cite{eroms2009,giesbers2012,zhang2013} where the measured transport gap has been attributed to localization rather than resulting from a band gap.

We have also shown that the size of the zero-temperature transport gap of a disordered graphene antidot lattice is linearly proportional to the radius of the antidots and inversely proportional to the square of the antidot lattice periodicity. The computed charge carrier mobilities in disordered graphene antidot lattices agree well with experimentally measured values and have been found to be fairly insensitive to the strength of the geometrical disorder.

\begin{acknowledgments}
This research has been supported by the Academy of Finland through its Centres of Excellence Program (Project No. 251748) and the National Natural Science Foundation of China (Grant Nos. 11404033 and 51202032). We acknowledge the computational resources provided by Aalto Science-IT project and Finland's IT Center for Science (CSC).
\end{acknowledgments}

\end{document}